\documentclass[%
reprint,
superscriptaddress,
amsmath,amssymb,
aps,prl,
]{revtex4-2}

\usepackage{graphicx}
\usepackage{dcolumn}
\usepackage{bm}
\usepackage[hidelinks]{hyperref}

\usepackage{mathtools}
\usepackage[caption=false]{subfig}

\usepackage{color}

\newcommand{\ie}{{\it i.e.}}
\newcommand{\eg}{{\it e.g.}}

\begin{document}
	
	
	\title{Homophily based on few attributes can impede structural balance
	%
	%
	}
	
	\author{Piotr J. G{\'{o}}rski}
	\affiliation{Faculty of Physics, Warsaw University of Technology, Koszykowa 75, PL 00-662 Warsaw, Poland}
	\author{Klavdiya Bochenina}%
	\affiliation{%
		ITMO University, Kronverkskiy av. 49, RU 197101 Saint Petersburg, Russia
	}%
	
	\author{Janusz A. Ho{\l}yst}
	\affiliation{Faculty of Physics, Warsaw University of Technology, Koszykowa 75, PL 00-662 Warsaw, Poland}
	\affiliation{%
		ITMO University, Kronverkskiy av. 49, RU 197101 Saint Petersburg, Russia
	}%
	\author{Raissa M. D'Souza}
	\affiliation{%
		Department  of  Mechanical  and  Aeronautical  Engineering,  University  of  California,  Davis, 95616, USA
	}%
	\affiliation{%
		Santa Fe Institute, Santa Fe, 87501, USA
	}%
	
	\date{\today}
	
	\begin{abstract}
        
        Two complementary mechanisms are thought to shape social groups: 
        homophily between agents and structural balance in connected triads. 
        Here we consider $N$ fully connected agents, where each agent has $G$ underlying attributes, and the similarity between agents in attribute space (\ie, homophily) is used to determine the link weight between them. 
  %
%
        To incorporate structural balance we use a triad-updating rule where only one attribute of one agent is changed intentionally in each update, but this also leads to  
        accidental changes in link weights and 
        even link polarities. 
        The link weight dynamics in the limit of large $G$ is described by a Fokker-Planck equation 
        from which the conditions for 
        a phase transition to a fully balanced state with all links positive can be obtained.  This ``paradise state" of global cooperation is, however, difficult to achieve requiring $G > O(N^2)$ and $p>0.5$, where the parameter $p$ captures a  willingness to consensus. 
        Allowing edge weights to be a consequence of attributes naturally captures homophily and reveals 
        %
        that many real-world social systems would have a subcritical number of attributes necessary to achieve structural balance. 
                
             
	\end{abstract}
	
	\maketitle
	
	
	\section{\label{sec:level1}Introduction} 
	
Recent experiments and analysis show that social networks are influenced by a set of distinct processes with possibly competing interactions \cite{Yap2015}. 
Two important processes are structural balance and homophilic  relations between agents. 
Structural balance (also called Heider balance or social balance) 
considers links 
between a set of three agents 
and assumes that recurring 
social interactions 
would lead towards eliminating tensions between them, eventually leading to a balanced triad~\cite{heider, bonacich, harary, kk, antal3, Marvel2009, strogatz, Zheng2015, Belaza2017, Gorski2017}. 
Using this principle, Antal {\it et al}. \cite{antal3} introduced a now seminal 
link-evolution model of local triad dynamics (LTD).
They showed the existence of a phase transition from a quasi-stationary state to a  ``paradise" state (with full structural-balance and no negative links). 
The LTD model is formulated at the level of link polarity (\ie, whether the link weight is positive or negative).
%
We consider a more basic notion, that agents have a set of underlying attributes which give rise to the link weights 
and that these co-evolve towards 
structural balance. 
  
Existing coevolution models \cite{Chen2014,Parravano2016,Singh2016,Lee2016,DU2016,Deng2016,Saeedian2017,Gao2018} study reaching the states of {polarisation} and {segregation} in terms of agents' interactions and opinions by assigning scalar (usually binary) polarities to links and at most two-dimensional attributes to nodes.
%
Instead, following the Axelrod model of culture dissemination \cite{Axelrod1997,Castellano2009,Deffuant2000}, we assume that each of $N$ agents possesses $G$ categorical Boolean attributes
and that 
edge weights decrease with distance between agents in the $G$-dimensional attribute space. 
We introduce an attribute-based local triad dynamics (ABLTD) model, where rather than changing the polarity of a link in attempt to increase structural balance, a more fine-grained change is made and one underlying attribute of one agent is changed.
In the large $G$ and $N$ limit the link weight dynamics can be described by a Fokker-Plank equation from which
we show that 
the phase transition observed by Antal {\it et al}. occurs for $G>O(N^2)$. 
For most real systems, the number of known attributes would be subcritical and the system would not be able to achieve full balance and global cooperation. 
Allowing the edge weights to depend on underlying attributes 
captures the perspective that the relations between people are dependent on the people themselves \cite{Rambaran2015,Nigam2018}.
Moreover, it lets us rigorously unify the principles of homophily and of social balance and analyze the thermodynamic limit. This reveals that homophily with only few attributes can prevent structural balance. 

	\section{\label{sec:model}Model}
	
	We consider a complete undirected signed network with no self-loops of $N$ agents labeled as  $i=1,2, \dots, N$. A state of each agent is described by  $G$-dimensional column vector of  categorical Boolean attributes  $\mathbf{A}_i=\{a_i^g\}$ where $g=1,2,...,G$ and $G$ is odd as in \cite{Axelrod1997,Castellano2009,Deffuant2000}.
	Each attribute $a_i^g$ is initially assigned $+1$ or $-1$ with equal probability.  
	The attributes correspond to agents' opinions or preferences about $G$ distinct subjects and they can change in time. 
	This allows a natural embedding in a Hamming space with $x_{ij}$ denoting the distance between agents $i$ and $j$ 
	and the polarity of their relation $P_{ij} = sgn(x_{ij}) \in \{ \pm 1\}$. 
	The polarities, $P_{ij}$, 
	depend on weights ($x_{ij}$) derived as a dot product: $x_{ij}= \dfrac{1}{2G}\mathbf{A}_i^T \cdot\mathbf{A}_j$. 
	It follows that $P_{ij}$ is positive (negative) if  more than half of the attributes of nodes $i$ and $j$ are the same (different).
	%
	
	
	\begin{figure}
		\centering
		\includegraphics[width=\linewidth]{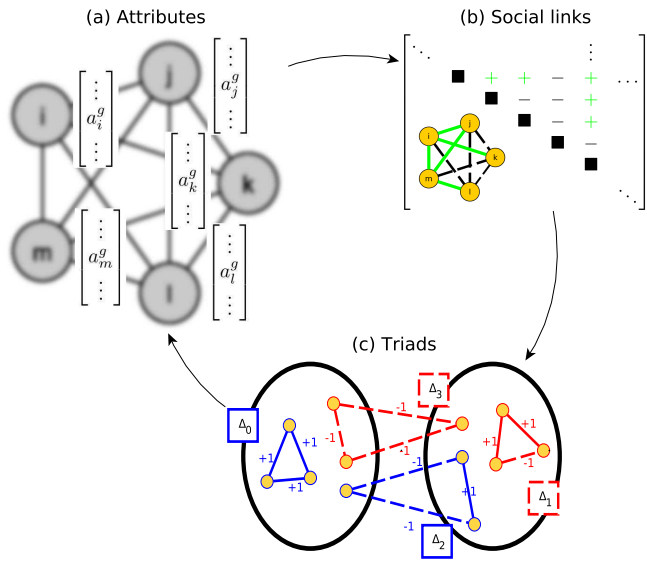}
		\caption{\label{fig:model} Coevolution of attributes, edge polarities and triadic relations. Nodes' attributes (a) determine the relations between agents (b) and formation of 
		balanced and unbalanced triads (c). 
		Unbalanced triads drive the evolution of node attributes (a) and so on... In (b) positive and negative links are denoted with {\color{green} +} and {\color{black}$-$}, respectively. Panel (c) presents the four types of possible triads: two balanced ($\Delta_0$ and $\Delta_2$) and two unbalanced ($\Delta_1$ and $\Delta_3$) denoted with blue and red colors, respectively. 
		If only $\Delta_0$ and $\Delta_2$ are present a group polarization is observed, \ie, agents can be divided into two hostile groups with all links within members of the given group positive and with cross-links negative. However, having some unbalanced triads such a division does not exist. 
}
	\end{figure}
	
	With this in place, we apply the 
	ABLTD dynamics. 
	In each step a random triad $\Delta(ijm)$
of nodes is picked. The triad is said to be {\it balanced} if the product of polarities of all its three links is positive, \ie,   $P_{ij} P_{jm}  P_{mi} =1$. Following  \cite{antal3}, all triads can be classified by a type $\Delta_k$ ($k=0,1,2,3$), 
where $k$ is the number of negative links in the triad. Unbalanced triads are of types $\Delta_1$ and $\Delta_3$. 
If  $\Delta(ijm)$ is unbalanced 
we flip one attribute (\eg, $a_i^g$) from one of nodes (\eg, $i$) so that the link weight 
changes by $\pm \dfrac{1}{G}$ towards the desired polarity. 
For a triad  $\Delta_3$ any link can be chosen since it is symmetric. For a triad $\Delta_1$ with probability $p$ a negative link is chosen and with probability $(1-p)$ one of the positive links is chosen. Note that $p$ thus captures the eagerness of the agents to achieve consensus. 
	%
	The coevolution of all three levels of the systems structure is depicted in Fig.~\ref{fig:model}.

	\section{\label{sec:level2}Asymptotic analytical solutions}
	
	
	Every state of the system can be defined by the specific placement of agents at points of Hamming space. Due to indistinguishability of agents many such states are equivalent. 
	For any given state one can in principle 
	calculate the transition probabilities to all other states 
	and with those probabilities 
	calculate exact measures of balanced states. 
	However, 
	the number of possible states grows very rapidly with $N$ and $G$ making the method infeasible in general. 
	
\begin{figure}
		\centering
		\includegraphics[width=1\linewidth]{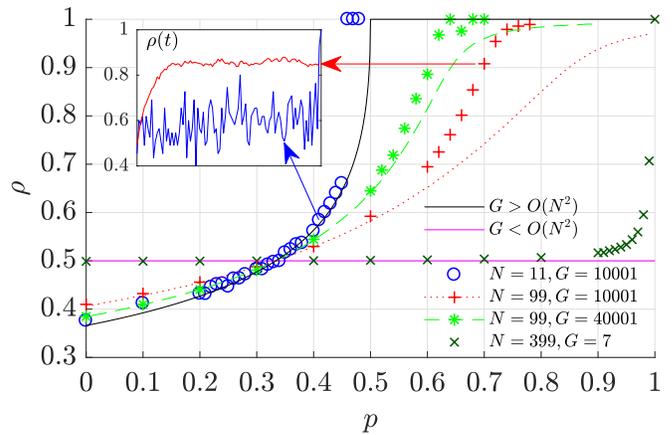}
		\caption{
			Steady states of the ABLTD model for different relations between numbers of attributes $G$ and nodes $N$. 
			Shown is the density of positive links $\rho$ as a function of probability $p$ (desire for consensus). When 
			$G>O(N^2)$ the system displays a transition to the ``paradise" state ($\rho=1$) provided $p$ is sufficiently large. 
			Analytical solutions (dashed {and solid} lines) fit the numerical results (markers) well for $p<0.5$.  Letting $G=O(N^\gamma)$ we see the fit is more accurate when $\gamma$ is larger. The solid black and magenta curves show the solutions in the asymptotic limit $N\rightarrow\infty$ when $\gamma>2$ and $\gamma<2$, respectively. The inset shows time evolution for two example systems. Both achieve a quasi-stationary phase, after which the smaller system ultimately reaches the paradise state. 
			Main plot shows the time and ensemble average of the link density when the system is in the quasi-stationary state if such a state is observed. 
		}
		\label{fig:fpe}
	\end{figure}

	For {numerical simulations of} a sufficiently large network after a number of updates one can observe the following states: (a) a stationary, paradise solution where only triads $\Delta_0$ exist, (b) a stationary, non-paradise solution where balanced triads $\Delta_0$ and $\Delta_2$ exist, or (c) a quasi-stationary solution where all types of triads coexist. 
	Networks of type (c) are observed frequently during simulations for many different parameter values. Such networks are unbalanced yet 
can be characterized by the approximately constant values of average measures such as positive link density $\rho$ or triad densities (see the inset of Fig. \ref{fig:fpe}). Fluctuations of these measures are visible 
and these  fluctuations lead finite-sized networks to ultimately reach one of the balanced solutions (a) or (b). 
The size of these fluctuations decrease with $N$. With $N\rightarrow \infty$ the fluctuations vanish, meaning that in the thermodynamical limit a system will stay in a quasi-stationary state never reaching a balanced solution. 
	
   We assume  $N$ and $G$ are both large numbers. 
	%
The possible values of the edge weights $x_{ij}$ (from now on denoted simply as $x$) form a discrete set: 
 $\mathbb{D}=\{-0.5, -0.5+1/G, -0.5+2/G, ..., 0.5\}$. 
	Any change in link weight is in increments $\pm1/G$. 
	Consider a random walk on $\mathbb{D}$. Let us denote `jumping right' as changing the weight to be closer to 0.5 and `jumping left' as changing the weight to be closer to $-0.5$ with 
	respective probabilities denoted by $r$ and $l$.
	During an update a walker stays in place with probability $1-(r+l)$. 
	There are two possible reasons  of each jump: intentional and accidental. The former is related to collective  social interactions  pushing triads towards structural balance (\ie, moving link weights towards 0). 
An accidental change (AC) is related to the fact that an intentional change (IC) in one triad triggers $N-3$ different ACs in other triads. An AC may result in a jump in either direction but with different probabilities dependent on the weight $x$. 
	The jump probabilities can be easily calculated. For instance, probability of `moving right' can be written as: $r\equiv P(r)=P(r|AC)P(AC)+P(r|IC)P(IC)$. 
	%
	%
	%
	The probabilities $r$ and $l$ are dependent on the weight $x$ (see \cite{SM}) as:
	\begin{align}
	r(x)  & =  \left\{\begin{matrix}
	(0.5-x)a_+ & \text{for } x>0 \\ 
	(0.5-x)a_-+i_- & \text{for }x<0
	\end{matrix}\right.\\
	l(x)  & =\left\{\begin{matrix}
	(0.5+x)a_++i_+ & \text{for } x>0 \\ 
	(0.5+x)a_-& \text{for }x<0
	\end{matrix}\right.,
	\end{align}
	where coefficients $a_\pm$ and $i_\pm$ represent probabilities $P(AC|x\gtrless~0)$ and $P(IC|x\gtrless 0)$, respectively and  in the mean-field approximation  can be calculated as functions  of density of positive links $\rho$ and model parameters $N$ and $p$ (see \cite{SM}). 
	
	Assuming an infinite number of possible states ($G\rightarrow\infty$) one can describe the evolution of the probability that a link has a weight $x$, denoted $W(x,t)$, using the Fokker-Planck equation (FPE), 
	
	\begin{equation}
\dfrac{\partial W(x,t)}{\partial t}=\Big[-\dfrac{\partial}{\partial x} c(x)+\dfrac{\partial^2}{\partial x^2}D(x)\Big]W(x,t),
\label{eq:fpe}
\end{equation}
with drift $c(x)\propto r-l$ and diffusion $D(x)\propto r+l$ \cite{ben2008fundamental,Codling2008}. 
The quasi-stationary solution of (\ref{eq:fpe}) can be expressed  by the corresponding potential as $W_{st}(x) \propto e^{-\phi(x)}$, where 
	
\begin{equation}
	\phi(x)=\dfrac{2G}{a_\pm+i_\pm}\Big(a_\pm x^2+i_\pm|x|\Big).
	\end{equation}
	This allows us to derive an analytical equation for the quasi-stationary values of positive link density $\rho$, as: $\rho={\int_{x>0} W_{st}(x)dx  }$. 
	The right hand side transforms into the transcendental function containing, among others, the cumulative standard normal distribution $\Phi$ dependent on density $\rho$, probability $p$ and the ratio of parameters $\dfrac{G}{N^2}$, see \cite{SM} for details. 
	The comparison of this analytical solution and numerical results for networks of $N=11$, $99$ and $399$ is shown in Fig. \ref{fig:fpe}. The fit is better for $p<0.5$ and when $\dfrac{G}{N^2}$ is larger. 
	One can observe a phase transition  displayed by  the  density of negative links $\rho$ that approaches one when the parameter $p$ crosses a critical value  dependent on $N$ and $G$.    
	
	\begin{figure}
		\centering
		\includegraphics[width=1\linewidth]{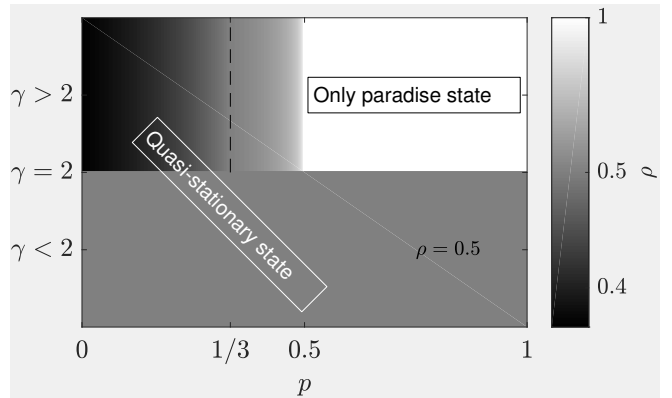}
		\caption{
			Phase diagram for the ABLTD model in the asymptotic limit. Axes represent probability $p$ and exponent $\gamma$, which describe agents' willingness to consensus and relation between numbers of nodes $N$ and attributes $G$ ($G=O(N^\gamma)$), respectively. The brightness represents the expected density of positive links $\rho$. When $\gamma>2$ and $p\ge0.5$ the only possible state is a paradise state ($\rho=1$). Otherwise the system reaches the 
			quasi-stationary state. For $\gamma<2$ for all $p$ the densities of positive and negative links are equal ($\rho=0.5$) {as are the number of balanced and unbalanced triads}. For $\gamma>2$ such equality is obtained for $p=1/3$ (marked by dashed line). 
		}
		\label{fig:phase}
	\end{figure}
	
	
		When $N\rightarrow\infty$ the density $\rho$   is dependent on the relation between $N$ and $G$ as a solution of equation
	
	\begin{equation}
	\rho=\dfrac{\Phi_+}{\Phi_++\exp{\Big(\dfrac{G}{N^2}\dfrac{C_-^2-C_+^2}{2}\Big)}\Phi_-},
	\label{eq:rho}
	\end{equation}
	where $\Phi_\pm=\Phi\Big(-C_\pm\dfrac{\sqrt{G}}{N}\Big)$
and $C_{\pm}$ are rational functions of $\rho$ and $p$ (see \cite{SM}).  	
 
To quantify the study, let us assume 
$G=O(N^\gamma)$. 
	\\\noindent
	$\bullet$
	If $\gamma>2$, then using L'Hospital's rule we obtain $\rho=\dfrac{C_-}{C_++C_-}$, which can be transformed into: 
	\begin{equation}
	(\rho-1)\Big((6p-2)\rho^2-2\rho+1\Big)=0.
	\end{equation}
	The solutions are the paradise state $\rho=1$ and a quasi-stationary solution $\rho=(1+\sqrt{3(1-2p)})^{-1}$ (the black line in Fig.~\ref{fig:fpe}). 
	Thus for $\gamma>2$ the ABLTD model is equivalent to the LTD model where a phase transition is observed at  $p=0.5$. 
	Below $p<0.5$  the system fluctuates around a quasi-stationary state. For $p>0.5$, 
		the system ends up in 
		the structurally balanced paradise state.
	
	\noindent$\bullet$
	If $\gamma<2$, then Eq (\ref{eq:rho}) transforms into simply $\rho=0.5$, where the quasi-stationary state with an equal number of positive and negative links exists. (The magenta line in Fig.~\ref{fig:fpe}.) It can be also shown that in such a state the numbers of balanced and unbalanced triads are the same. 
	
		The system phase diagram as obtained in the asymptotic limit $N\rightarrow\infty$, is presented in Fig. \ref{fig:phase}.
		Numerical simulations show that one does not need a large number of nodes to obtain a qualitatively similar transition. For instance, 
		see the results for $N=11$
		(the blue dots) in Fig.~\ref{fig:fpe}. 
		For such small systems sizes quasi-stationary fluctuations in $\rho$ may be difficult to observe
		and the transition occurs for values $p<0.5$.

\begin{figure}
		\centering
		\subfloat{\begin{tabular}{cc}
				\includegraphics[width=0.5714\linewidth]{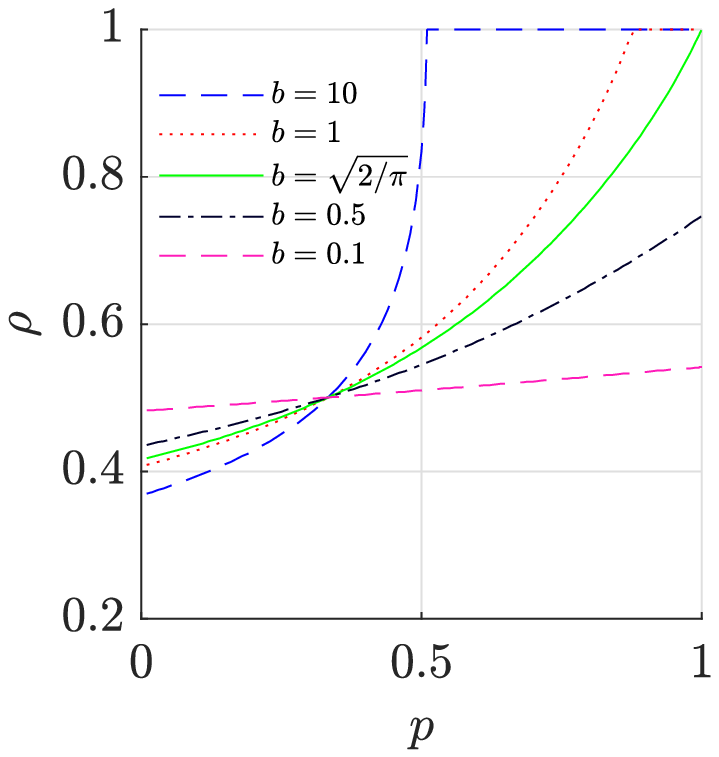} &\includegraphics[width=0.4286\linewidth]{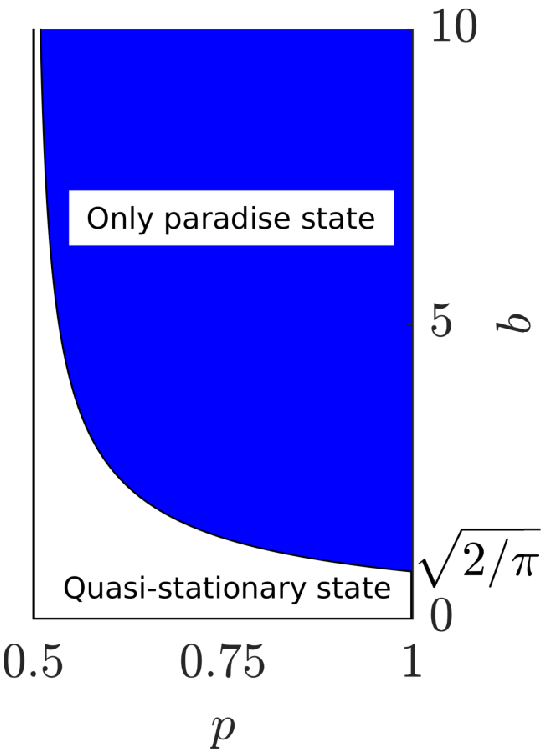} \\
				(a) & (b) 
			\end{tabular}}
			\caption{
				For the case $G=(bN)^2$ the existence of a phase transition in ABLTD 
				depends on $b$. 
				Here, the $N\rightarrow\infty$ limit is analyzed. 
				Panel (a) shows the density of positive links $\rho$ for quasi-stationary solutions. For $b>\sqrt{\dfrac{2}{\pi}}$ there exists a critical point $p^*$ such that for $p>p^*$ the only solution is paradise ($\rho=1$). 
				For $b<\sqrt{\dfrac{2}{\pi}}$ paradise cannot be reached for any $p$. 
				Panel (b) shows the phase diagram for such a system. The phase boundary is given by the relation $p^*(b)$. 
			}
			\label{fig:fpe_ng2}
		\end{figure}

	\noindent$\bullet$
	If $\gamma=2$, then the system reaches an intermediate asymptotic solution (see Fig.~\ref{fig:fpe_ng2}a). 
	Assuming $G=(bN)^2$ we obtain the equation for the critical point $p^*$, above which the quasi-stationary state disappears:
	\begin{equation}
	\dfrac{1}{\sqrt{2\pi}}\exp{\Big(-\dfrac{1}{2}b^2(1-p^*)^2\Big)}=bp^*\Phi{\Big(-b(1-p^*)\Big)}
	\label{eq:ps:a}
	\end{equation}
%
The phase transition to the paradise state exists so long as $b\ge\sqrt{\dfrac{2}{\pi}}$. The relation $b(p^*)$ is the phase boundary plotted in Fig. \ref{fig:fpe_ng2}b.

		\section{No transition in 3-node network} 
		
		Let us also explore the simplest case, \ie, the network comprising only one triad. For such a system with small number of attributes it is possible to calculate exactly the probabilities of stationary balanced states. With higher $G$,  one needs numerical simulations, and with $G\rightarrow\infty$ analytical calculations are again possible. 
		In such a case 
		and assigning {attributes at random,} the initial distribution of triads of types $\Delta_0$, $\Delta_1$, $\Delta_2$ and $\Delta_3$ is equal to $\dfrac{1}{8}$, $\dfrac{3}{8}$, $\dfrac{3}{8}$ and $\dfrac{1}{8}$. 
		Still, the attribute update causes two links to change their weights, but since $G\rightarrow\infty$ then probability of crossing the polarity change threshold (\ie, $x=0$) for both of the links at the same time is negligible.
		Thus, with one edge flip a triad of type $\Delta_3$ always turns into $\Delta_2$. The fate of a triad of type $\Delta_1$ depends on the parameter $p$. 
		Weights' evolution for links in such a triad is given as follows:
		\begin{align}
		\dot{x}_{-}&=p-(1-p)x_- 
		\label{eq:N3-diff1}\\ 
		\dot{x}_{+}&=-\dfrac{1-p}{2}-\dfrac{1+p}{2}x_+,
		\label{eq:N3-diff2}
		\end{align}
		where $x_\pm$ represents the weight of one of the two positive links or a negative link in this triad. 
		Additive constant terms in Eqs (\ref{eq:N3-diff1}-\ref{eq:N3-diff2}) are related to the incidental change of a corresponding link, while the linear parts to the accidental change of an adjacent link. 
		These equations are valid up to the moment when any of the link weights cross 0, which is equivalent to the flip of the link polarity. 
		%
		%
		
\begin{figure}
			\centering
			\subfloat{\begin{tabular}{cc}
					\includegraphics[width=0.3846\linewidth]{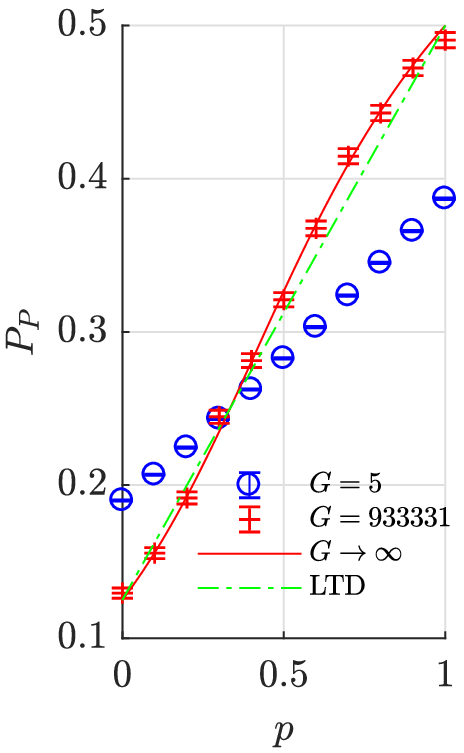} &\includegraphics[width=0.6154\linewidth]{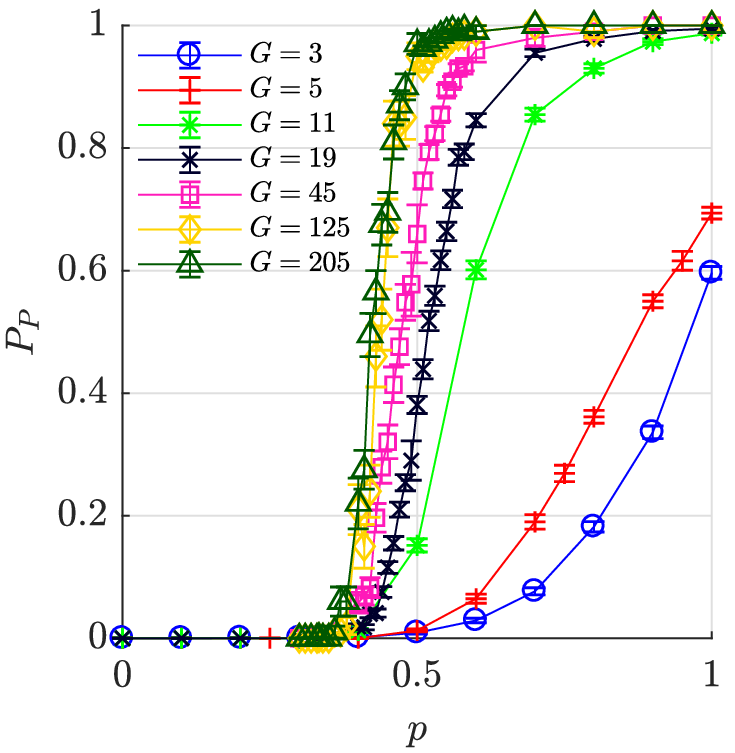} \\
					(a) & (b) 
				\end{tabular}}
			\caption{ 
				Observed transitions for small systems of size: (a) $N=3$ and (b) $N=11$. Plots show probability of reaching paradise $P_P$ in ABLTD as a function of incentive to achieve consensus $p$. 
				(a)
				In the case of only one-triad systems there is no transition. The figure compares $P_P$ for the numerical results of two diverse numbers of attributes $G$ and the approximate $P_P(p)$ in the limit $G\rightarrow\infty$. Numerical results for large $G$ agree with the analytical solution. The expected 
				{analytical} results for the LTD model are also shown for comparison. 
				(b)
				In the system with $N=11$ the transition is observed and becomes sharper with increasing $G$. 
			}
			\label{fig:small-systems}
		\end{figure}

		Probability of a change $\Delta_1\rightarrow\Delta_0$ is equivalent to calculating the probability that a negative link will be first to change polarity. 
Equations (\ref{eq:N3-diff1}-\ref{eq:N3-diff2}) let us calculate $P_{\Delta_1\rightarrow\Delta_0}$ (see \cite{SM} for details) 
		and derive an asymptotic probability of reaching the $\rho=1$ paradise state: $P_P(p)=\dfrac{1}{8}+P_{\Delta_1\rightarrow\Delta_0}(p)\dfrac{3}{8}$. 
		The results for numerical simulations averaged over initial conditions and an asymptotic solution as well as the results for the LTD model are presented in Fig. \ref{fig:small-systems}a. 
		These solutions indicate that for the smallest system, no matter how large the incentive to cooperate $p$ is, $P_P$ can never be 1. 
		For larger systems $P_P$ can show a sharp increase, as observed in the $N=11$ example shown in Fig. \ref{fig:small-systems}b. 
However, in such a small network a large number of attributes are needed. Otherwise, for instance for $G=3$ or $G=5$ paradise may not be likely for any value of $p$. 
		
		
		\section{Invariant features}
		
		When $p=1/3$ different measures (\eg, probability of reaching paradise $P_P$, 
		or mean positive link density $\rho$ for both balanced and quasi-stationary states) are independent of the number of the attributes $G$. 
		%
		This observation has been confirmed in all numerical and analytical results for all network sizes. 
		It is related to the fact that for $p=1/3$ for all unbalanced triads all nodes and links are updated with the same probability. 
		For this $p$ the exact values for any $N$ can be calculated only for some special cases, \eg, for $G=1$: $\rho(G=1)=0.5$ and $P_P(G=1)=2^{-(N-1)}$. We have statistically confirmed that above relations are valid for larger $G$ (see \cite{SM}).

		\section{Conclusions}

We developed a rigorous framework that unifies the principles of homophily and structural balance and which is amenable to thermodynamic analysis. Rather than most previous works on structural balance, 
{which manipulate the polarity of links}, 
we consider that link weights are a consequence of underlying attributes of nodes. This captures the fundamental perspective that agents have preferences, and also embeds our system in a Hamming space giving a quantitative measure of similarity between agents. Extending the LTD link-evolution model to the more nuanced ABLTD attribute-evolution model 
reveals the interplay of homophily and structural balance. 
        A phase transition from  a quasi-stationary state to a paradise state can be observed and 
        the nature of the 
        transition depends on the exponent $\gamma$ 
        relating the number of attributes $G$ and nodes $N$, where $G\sim N^\gamma$. 
		Our study  emphasizes the importance of attributes for structural balance theory and shows that homophily can impede a system from achieving balance and global cooperation. 

Our analytic and numerical results indicate that the LTD and ABLTD models are equivalent in the limit $N,G\rightarrow\infty$ and $G>O(N^2)$. In such a case below $p<0.5$ the system fluctuates around a quasi-stationary state. For $p>0.5$ 
the system achieves the 
paradise state. When $G\le O(N^2)$ the asymptotic results are different. For instance when $G<O(N^2)$, then the quasi-stationary state with equal number of balanced and unbalanced triads exists for all $p$.
		
		
		
An experimental analysis of structural balance would require proper datasets of social relations between agents. Such datasets should comprise time-varying data not only of agents' relations but also of their opinions and personal characteristics. 
Knowing both the relations and attributes one can estimate the possible fates of the system. 
As an example let us take a small group of agents. Our model indicates that paradise is unlikely in such a situation even with high eagerness towards consensus (parameter $p$). 
For a team (sports or industry) a common goal (a strong single attribute) may be sufficient in achieving a paradise state but in other cases a leader needs other interventions to create bonding relations and prevent division into separate subgroups, especially when the number of attributes is large~\cite{WaagenPRE2015}. 
Results from our model can explain why 
structural balance 
is not observed in some experiments over large social networks \cite{Leskovec2010a,Leskovec2010,Szell2010}. It may be the outcome from internal features of the network itself. 
The state without balance is also the natural state, as it is shown in our model for large number of nodes or attributes. 
		

		Our work reveals the competing tension between homophily and structural balance and the importance of accounting for node attributes, especially when the number of the attributes is small, which we believe is the most 
		common real-world scenario. 
		
				
		
		\begin{acknowledgments}
			This research has received funding as RENOIR Project from the European Union’s Horizon 2020 research and innovation programme under the Marie Sk{\l}odowska-Curie grant agreement No. 691152, by Ministry of Science and Higher Education (Poland), grant Nos. W34/H2020/2016, 329025/PnH/2016,  by National Science Centre, Poland Grant No. 2015/19/B/ST6/02612, and by the U.S. Army Research Office MURI award W911NF-13-1-0340. K.B. and J.A.H. were partially supported by the Russian Science Foundation, Agreement \#17-71-30029 with co-financing of Bank Saint Petersburg. 
		\end{acknowledgments}

		

%
		
	\end{document}